**Authors**

Jacob D. Beckmann (corresponding author), Kosta Popovic

**Title**

Combination of multiple neural networks using transfer learning and extensive geometric data augmentation for assessing cellularity scores in histopathology images

**Affiliations and Addresses**

J. Beckmann was with the Physics and Optical Engineering Department at Rose-Hulman Institute of Technology, Terre Haute, IN 47803 USA (e-mail: [beckmaj1@rose-hulman.edu](mailto:beckmaj1@rose-hulman.edu)). He is currently with Honeywell FM&T

K. Popovic is with the Physics and Optical Engineering Department at Rose-Hulman Institute of Technology, Terre Haute, IN 47803 USA (e-mail: popovic@rose-hulman.edu, phone: 812-877-8190, fax: 812-877-8309)

**Declarations**

**Ethics approval**

   No approval of ethics was required for this work.

**Consent to participate**

   No consent for participation was required for this work.

**Consent to publish**

   No consent for publishing was required for this work.

# Combination of multiple neural networks using transfer learning and extensive geometric data augmentation for assessing cellularity scores in histopathology images

*Abstract* — Classification of cancer cellularity within tissue samples is currently a manual process performed by pathologists. This process of correctly determining cancer cellularity can be time intensive. Deep Learning (DL) techniques in particular have become increasingly more popular for this purpose, due to the accuracy and performance they exhibit, which can be comparable to the pathologists. This work investigates the capabilities of two DL approaches to assess cancer cellularity in whole slide images (WSI) in the SPIE-AAPM-NCI BreastPathQ challenge dataset. The effects of training on augmented data via rotations, and combinations of multiple architectures into a single network were analyzed using a modified Kendall Tau-b prediction probability metric known as the average prediction probability PK. A deep, transfer learned, Convolutional Neural Network (CNN) InceptionV3 was used as a baseline, achieving an average PK value of 0.884, showing improvement from the average PK value of 0.83 achieved by pathologists. The network was then trained on additional training datasets which were rotated between 1 and 360 degrees, which saw a peak increase of PK up to 4.2%. An additional architecture consisting of the InceptionV3 network and VGG16, a shallow, transfer learned CNN, was combined in a parallel architecture. This parallel architecture achieved a baseline average PK value of 0.907, a statistically significantly improvement over either of the architectures' performances separately (p<0.0001 by unpaired t-test).

There is no external sponsor or financial support to disclose.

ACKNOWLEDGMENT

Data used in this research was acquired from Sunnybrook Health Sciences Centre with funding from the Canadian Cancer Society and was made available for the BreastPathQ challenge, sponsored by the SPIE, NCI/NIH, AAPM, Sunnybrook Research Institute

## I. Introduction

Cancer is currently one of the most common diseases and killers in the world [1]. As a result, many resources have been invested towards developing new ways to treat the disease, as well as monitor recovery and treatment plans over time. One development used to track the progression of cancer over time is by quantifying the residual cancer burden. This method analyzes the malignant tumor tissue in order to compare the amount of cancerous nuclei present. The analysis of these tissue samples is currently left to trained medical professionals, resulting in a manual, time consuming, and sometimes subjective process. First the primary tumor bed must be identified from the tissue sample. This tumor bed is then split into different slides or smaller subsections. After the preparation for analysis, the assessment of cancer cellularity is recorded in increments of 10%, with occasional inclusion of an additional 5% or 1% tolerance [2]. For this reason, the current process of analyzing cancer cellularity and residual cancer burden can be a difficult task to perform quickly and accurately.

Current developments in the field of computer vision have helped to reduce the time of classifying various diseases from patients' images. Specifically, the application of Deep Learning (DL) has created a presence in the field of medical image analysis due to the ability to recognize patterns not easily identifiable by humans [3]. Convolutional Neural Networks (CNN) are a common form of DL architecture that are used for medical image analysis. CNNs are constructed of multiple layers containing filters to extract features such as curves and edges from image data.

Although CNNs have improved the speed of analyzing medical images, while also performing to pathologist standard in certain cases, the applications of these networks remain restricted, mainly due to the limited amount of readily available and well-labeled data for networks to train on. CNN performance can be greatly improved with the addition of appropriate training data [4]. A large and diverse, well-labeled dataset gives the network more features and patterns to consider during training, and creates a more robust network.

Additionally, many high performing CNNs are ones constructed of pre-trained networks as opposed to building ones from scratch [5]. While it is commonly accepted that building a CNN from scratch for a particular purpose provides the user with a more powerful tool, this process is not trivial. The initial construction of a new CNN architecture requires a strong mathematical background from the user, knowledge in filtering and feature extraction, and a strong understanding of neural network connections to achieve performance comparable to humans. The practice of transfer learning, or using pre-trained networks, has reduced the amount of background knowledge required, while maintaining networks' performance compared to trained medical professionals [6].

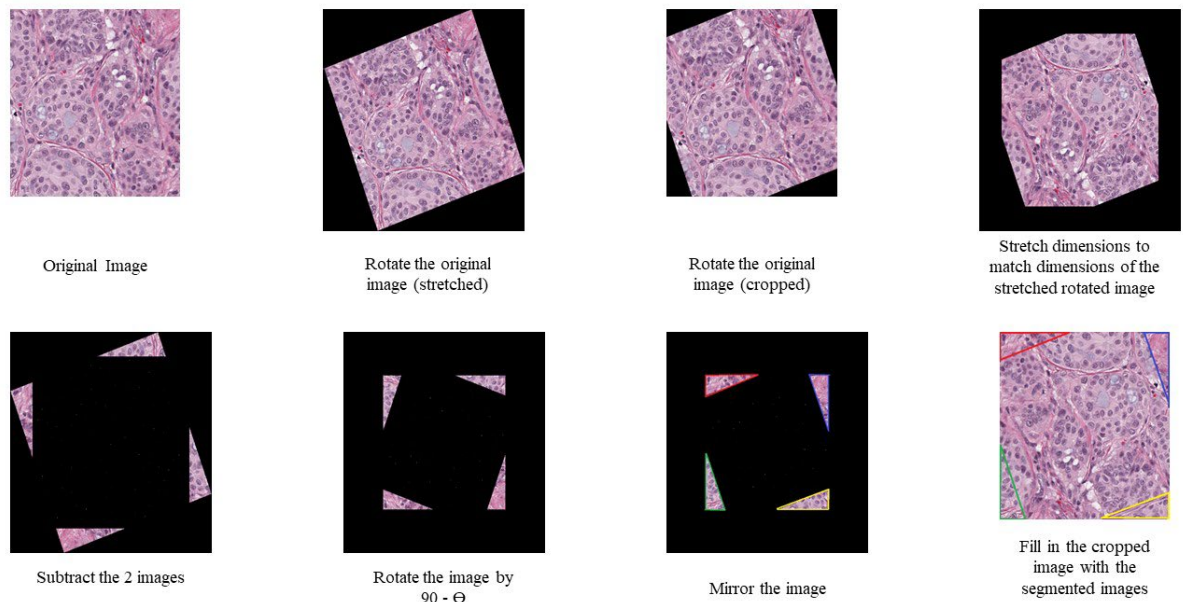

**Fig. 1** Proposed lossless rotation data augmentation technique

## II.  METHODS

The focus of this work is to analyze how the architecture of a CNN, and the amount of data used for training, affects the performance of ranking cancer cellularity within patches of whole slide images (WSI) from the SPIE-AAPM-NCI BreastPathQ [7] challenge dataset [8].

The first approach tested in this work proposed a form of data augmentation to alleviate the common issue of limited datasets. When training a neural network, it is important to ensure that the network has a large and diverse dataset to train on [4]. This data is used by the network, by learning which features and patterns within the data are essential over time, to make accurate predictions on future new data. However, in many cases, especially in the medical field, the amount of readily available data for a neural network to train on is limited. Additionally, some of the classes may be underrepresented due to poor statistics or low frequency of occurrence of a specific morbidity. To improve the range of features and patterns a network will learn, more diverse images are essential within a training set. The proposed data augmentation technique provides a novel form of lossless image rotation, capable of increasing a dataset by a factor of 360.

The second approach, dealing with CNN architecture, combines two transfer-learned networks of different sizes.

Transfer learning is a common technique used in the field of deep learning, due to the reduced need for well labeled data, construction time, and computational resources [9]. There are multiple types of transfer learning networks that can provide different results based on their filtering and number layers, but they are generally used individually. This work demonstrates the impact of constructing a network that can potentially extract more information by using complementary filters from two intrinsically different networks.

### A.  Data Augmentation

Data augmentation is a common technique to increase the size of a dataset. In the case of images, data augmentation consists of applying some modifications to the original images that retains all, or parts of the original information, but changes the image sufficiently in order to have the network analyze the image as new data.

Common forms of data augmentation consist of rotations or mirroring of an image. An image of cells that is rotated 90 degrees still contains the original cells but appears visually different. As a human, this insight comes from experience of having learned the relations and characteristics of cells and the properties of rotation over time. The network, however, does not have that benefit, and thus will need to train to draw the same conclusion. These methods can be applied to all images within a dataset to increase the number of images that a network can be trained on.

Although the generation of artificial data can increase the performance of a network, in some applications it is beneficial if the augmentation does not affect the information present in the original image. One example of augmentation that may lose information would be magnification. Although more details could be extracted in the area where the zoom occurs, information along the edges and other borders will be lost, and the network will be unable to consider that information. Augmentation techniques that provide some change in information compared to the original image are not considered

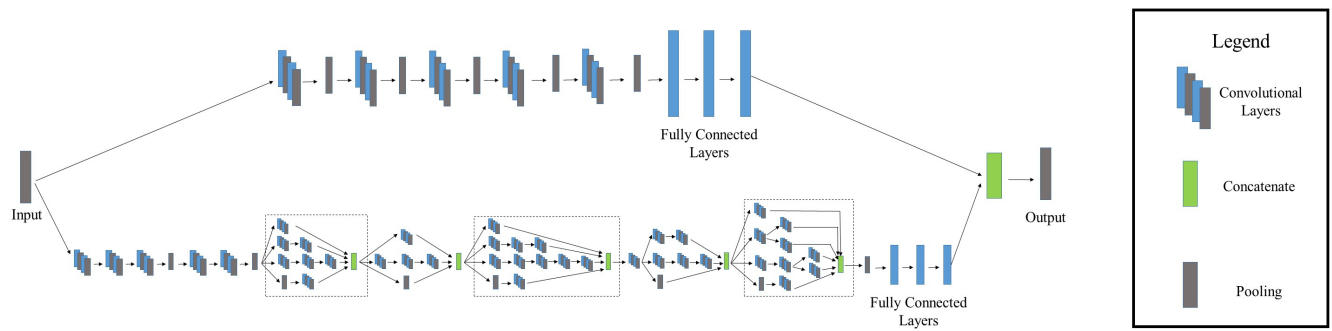

**Fig. 2** Visualization of the parallel architecture composed of VGG16 (Top) and InceptionV3 (Bottom)

in this discussion but have also been shown to improve network's performance [10].

To alleviate issues related to limited training data and loss of information through augmentation, our proposed method combines augmentation of a cropped rotated image and a resized rotated image to retain all original information and preserve original image resolution. A rotation with a cropped fit and a rotation with a resized fit is performed. An image subtraction of the two rotations is performed, and the resulting pixels, which were lost in the resolution-preserving rotation, are replaced into the image, creating a lossless augmentation process, and a potential increase in the training dataset by a factor of 360 if image is rotated by one degree following this process. This process is shown in detail in Figure 1.

### B. Transfer Learning

The construction of the neural networks was done using Keras, a Python library and TensorFlow backend. Two transfer learned models were used individually to generate a baseline for the dataset. The two networks used, InceptionV3 and VGG16, vary in their size and number of layers. InceptionV3 [11] is a deep network containing 48 convolutional layers, while VGG16 [12] is a shallow network containing 13 convolutional layers. The learning rate and model architecture were determined through deliberate overfitting. This process was performed in order to validate that the network would be capable of fully learning the dataset. Once the optimal values were determined, these parameters were kept consistent through all training sessions.

InceptionV3 and VGG16 perform the same processes of visually processing and interpreting images, but vary in the features and information they extract in their filtering layers to make their final outputs. This is based on the network architecture, and specifically the number of convolutional layers, and number of filters within these convolutional layers. VGG16, a shallow network which contains fewer convolutional layers, extracts simple features such as edges and curves. InceptionV3, a deep network with over triple the number of convolutional layers, can extract more complex features from images due to the increased number of filters [13]. Combining these networks into a single architecture may allow for a robust network containing simple and complex filters, which is not achievable in a single network architecture [5].

To create a parallel architecture, each architecture was first constructed and trained individually. Both networks were trained and validated on the same dataset. With both networks constructed, the output layers of each were removed from the architecture, and the last fully connected layers were connected into a new output layer. Additionally, the input layer of each network was removed, and a new single input layer was connected to the first convolutional layer of each network, to complete the parallel architecture, as shown in Figure 2.

### C. Training

The training dataset used for establishing the baseline of each network contained the original training images from the SPIE-AAPM-NCI BreastPathQ dataset [8]. In addition, the common augmentations of each training image rotated by 90, 180, and 270 degrees were included, for a total of 8,619 training images for the independent networks, and 9,097 training images for the combined network. Due to the limited computational resources available, these images were reduced from their original resolution 512 x 512 x 3 to 256 x 256 x 3, as this has been shown to be a good trade-off while maintaining network performance [14]. The training images were shuffled before being presented to the networks for training. The validation set contained 957 images to monitor the network's training and ensure overfitting did not occur for the independent networks, and 479 images for the combined network. The testing set consisted of 1,119 images that were never seen by the network during training or validation.

To reduce training time, monitor training performance over time, and prevent resource exhaustion, the amount of rotated training datasets implemented into a single session was reduced to sets of 30 random rotations. This allowed correlation of network outputs as more augmented data was added to the training set. To ensure the network does not overfit, a validation set of 3,352 images for the independent networks, and 1,678 images for the combined network, was used to compare the loss values, and PK value of the network per training epoch. After training, the network weights were saved and re-loaded into a newly constructed network for the next training session.

Two regularization methods were implemented to ensure the network does not overfit. Dropout layers were added to both networks and were established during the fine-tuning phase of network construction. A dropout layer weighted at 80% was put between the last fully connected layer and output in the InceptionV3 architecture [15]. The VGG16 network had two dropout layers weighted at 80%, one between the second and the final fully connected layer, and one between the final fully connected layer and the output layer. Additionally, early stopping with a tolerance of 10 epochs was implemented for each individual network. The combined parallel model implemented early stopping with a tolerance of 5 epochs. Additional hyperparameters for each network can be seen in Table 1.

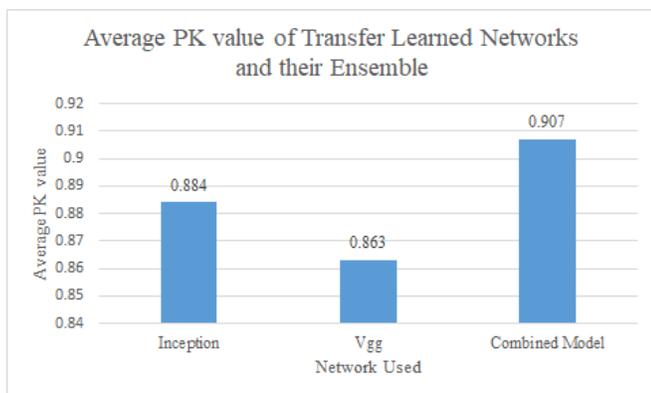

**Fig. 3** Average PK value of InceptionV3 and VGG16 Individually and in a Parallel Architecture

| Hyperparameters used for each network | | | |
|---|---|---|---|
| | InceptionV3 | VGG16 | Combined Model |
| Learning Rate | 1.00E-03 | 1.00E-05 | 1.00E-05 |
| Epochs | 2000 | 2000 | 50 |
| Batch Size | 16 | 16 | 14 |
| Early Stopping patience | 10 | 10 | 5 |
| Optimizer | Adam | Adam | Adam |
| Training/Validation Split | 80/20 | 80/20 | 95/5 |

**Table 1** Summary of hyper parameters used in each of the network configurations

## III. RESULTS AND ANALYSIS

The performance of the network was analyzed using a modified Kendall Tau-b ranking correlation, referred to as the average prediction probability PK [16]. This metric calculates a prediction probability between two pathologists and the submission, which was the official evaluation metric used for the SPIE-AAPM-NCI BreastPathQ challenge. The network was tested on the test data set provided by the BreastPathQ challenge organizers and scored via CSV submission [17]. The results of the baseline transfer-learned architecture performance and their ensemble can be seen in Figure 3. Testing InceptionV3 resulted in an average PK value of 0.884, while the VGG16 architecture achieved an average PK value of 0.863. The same process of establishing a baseline was repeated for the parallel architecture. The parallel architecture achieved an average PK value of 0.907, seeing an improvement in performance in comparison to InceptionV3 and VGG16 individually in ranking cancer cellularity in patches of WSI. The implementation of multiple transfer learned networks into a single architecture shows a significant improvement from either network used individually, in this case showing an improvement up to 2.5% from the InceptionV3 architecture and 4.9% from the VGG architecture. Unpaired t-test was run to compare the average PK values between the parallel architecture and the InceptionV3 architecture, which resulted in a statistically significant difference (p value was less than 0.0001).

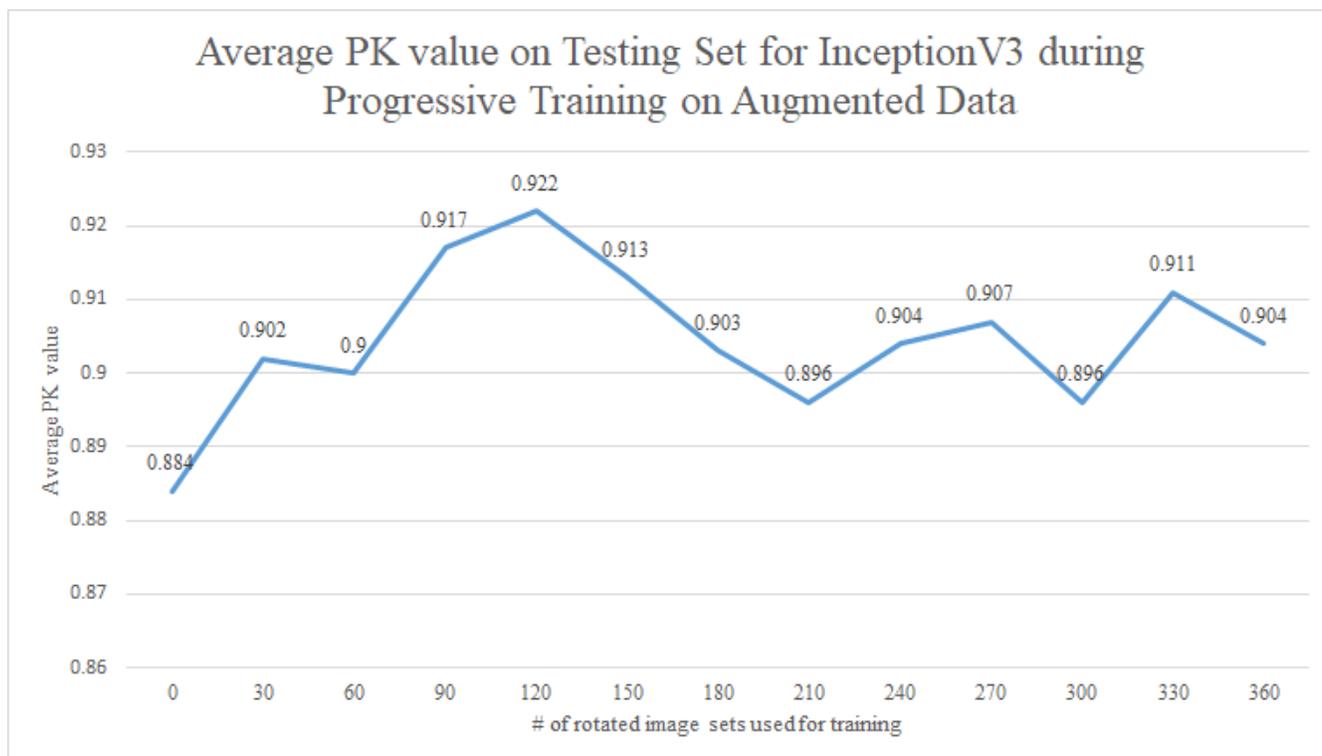

**Fig. 4** Progression of Average PK for InceptionV3 when trained on more augmented data

Further training on large amounts of augmented data provided variations in network output after every session, as shown in Figure 4. In all cases, InceptionV3 experienced an improvement in performance in comparison to the non-augmented image baseline with the best performing result at 120 rotations providing a 4.2% increase. However, notably the most obvious improvement occurred immediately after introducing the first 30 rotations (Figure 4), after which the average PK value hovered around this improved value for the rest of the tests. Nevertheless, even in the case where the improvement was the least (1.3%) compared to the baseline (at 210 rotations), the unpaired t-test resulted in a statistically significant difference between the two cases (p value was less than 0.0001). It is important to note even the addition of the first 30 rotations still adds value compared to baseline, as the baseline images included the common augmentations of each training image (rotated by 90, 180, and 270 degrees).

To gain a further understanding of how the InceptionV3 architecture parameters adjusted as more augmented data was presented, the filters from the trainable convolutional layers were extracted after 180 training iterations. A visualization technique [18] was used to extract a heatmap of the features extracted by a filter from a training data in the final convolutional layer, shown in Figure 5. The three filters shown demonstrate a more detailed feature extraction on three examples (in color) from the training data as the network trained on more augmented data. This demonstrates that the introduction of large amounts of data modifies the features being extracted by the network.

## IV. Discussion

Individually, the transfer learned networks provided results superior to trained pathologists when tasked to rank cancer cellularity within WSIs [19]. When it comes to individual performance, the deeper CNN, InceptionV3, outperformed VGG16. The combination of both the shallow and deep network outperformed both networks individually. The parallel architecture saw an increase from the highest individual architecture PK by 2.5%. The number of filters present within a single CNN architecture can improve the performance on the ability to rank cancer cellularity within a WSI. A statistical analysis was performed to further demonstrate that the test results provided by the parallel network are statistically significant compared to individual networks.

Providing the network with additional augmented data acquired via lossless image rotation demonstrates improved network performance. By training in sets of 30 different rotations, the performance was analyzed through a progression as shown in Figure 4. A statistical analysis was performed to demonstrate that the addition of the augmented data improves performance of the InceptionV3 network at statistically significant levels in every case compared to baseline.

## V. Conclusion

This work demonstrates how implementing large amounts of augmented data and combining multiple transfer learned networks into one architecture affects the ability of CNNs to rank cancer cellularity within patches of WSI. The baseline results demonstrate that InceptionV3, a deep transfer learned architecture with augmented data is capable of accurately ranking cancer cellularity with an average PK value of 0.884. Further, the work demonstrates that leveraging power from two intrinsically different transfer learned networks into a single architecture can improve the PK value to 0.907.

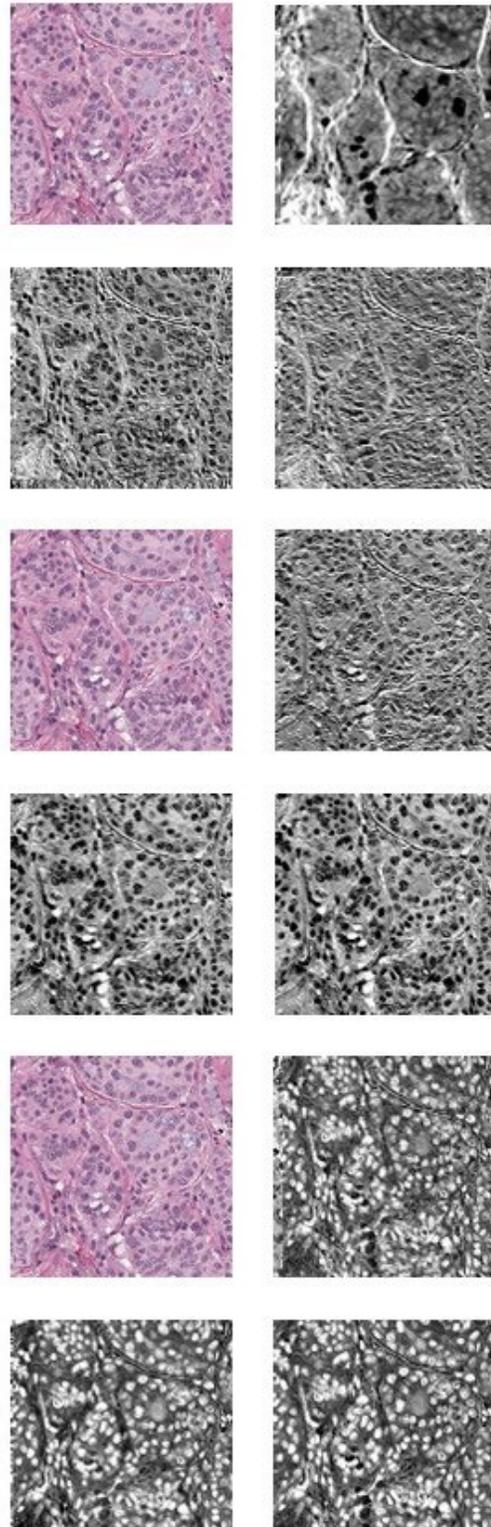

**Fig. 5** Visualization of three different convolutional filters applied to a WSI patch (top left) after training on no rotations (top right), 180 rotations (bottom left) and 360 rotations (bottom right)

A lossless data augmentation technique was proposed and provided an increase on the size of a training dataset by 360x the original size. This method provided improvement in average PK metric compared to the baseline for any amount of added data. Investigation of filters within the InceptionV3 network gives insight to the network's ability to train and extract new features (Figure 5).

Future work will investigate the proposed data augmentation technique as well as the parallel architecture performance with other datasets. Additionally, the investigation of network performance with these methods within a classification problem rather than a regression scenario will be sought.

CONFLICT OF INTEREST

On behalf of all authors, the corresponding author states that there is no conflict of interest.